# Designation of a binocular structure for complex sources of x-rays and neutron source

X.B. Yang[1], X.B. Qin[1,2a)], P.H. Liu[1], T. Li[1], K. Zhuang[1]

1 Beijing Engineering Research Center of Radiographic Techniques and Equipment, Institute of High Energy Physics, Chinese Academy of Sciences, Beijing 100049, China

2 X-lab, the Second Academy of China Aerospace Science and Industry Corporation Limited, Beijing, China

a)qinxb@ihep.ac.cn

Abstract

A structure, used as a complex source of X-rays and neutron source, is designed with the assistance of CST module tool. Particles of –H and electron, attracted from a plasma source, are designed to be separated by an energy selector. Soon after the separation, the trajectory and beam quality of -H is controlled by a special designation of potential barrier and potential well. And the acceleration of both particles is realized by three pairs of pierce similar pole to an energy of 150keV to produce x-rays/neutron. With proper optimization of structure and potential combination, a binocular structure with beam center distance of 15mm, beam diameter smaller than 3.5mm has been devised, which gives a feasible suggestion to produce two rays with one structure in the same time.

## 1 Introduction

X-rays and neutron source both play an important role in the domain of material physics, nondestructive test, archaeology, medical science, national defense, etc. Because of different interaction mechanism with atom, x-rays and neutron source show different characteristics when they interact with the same object. With the development of science and technology, the sole source cannot satisfy the higher need of the experiment [1-4]. The debut of complex source is realized by the combination of neutron source and x-rays equipment in 2012 at Missouri University of Science and Technology [5-7]. Ever since then, the development of this technique is pervasive among lots of institution around the world, and lots of impressive and exciting work has been done through this way [8-14]. What has to be pointed out is that, most of the experiment mentioned above is realized by combining large scale neutron experiment equipment such as, reactor and accelerator, with x-ray sources. This work comes up with a designation of a structure that can produce those two rays with one structure at the same time, based on the structure of neutron tube with negative hydrogen source [15-17]. The designed binocular structure of the complex source is mainly composed of three parts: the electron and negative hydrogen source, the separate apparatus, and the binocular accelerate structure.

## 2 Particle source

The particle source of –H and electron is designed to be a plasma source, but considering the absence of plasma source simulation module, a source of pierce gun is instead at the first designation, Fig1(a). During the simulation, the surface of the cathode is divided into 12 equal parts, Fig1(b), and each part releases electron or –H in sequence, with the purpose of emitting the two kinds of particles in the same time and defining the transverse outline of the beam.

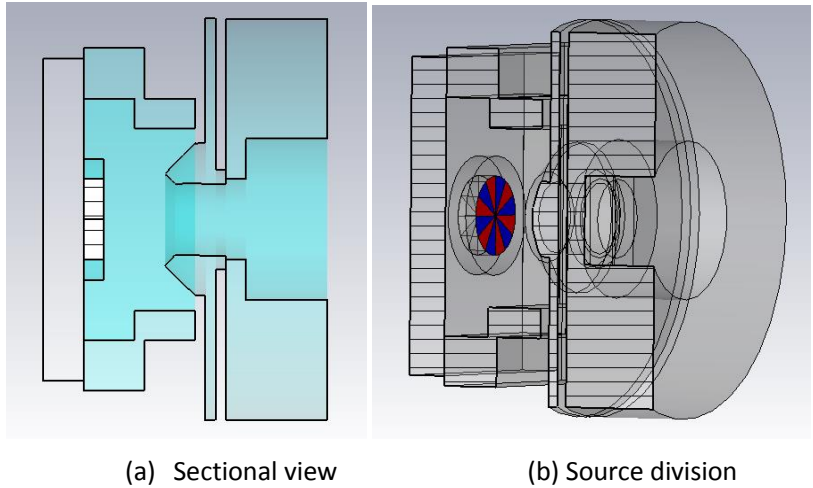

(a) Sectional view            (b) Source division

Fig1. (a) Sectional view of the particle source, with a similar structure of pierce gun, consists of cathode, focus and anode. (b) A circular surface on the cathode with a diameter of 4mm is divided equally into 12 parts, and each part emits electron or –H in sequence from the plane 0.2mm away from the cathode surface.

3 Rules of separation

Different with the static magnetic flipping field of neutron tube, during the designation of the separate apparatus of the binocular structure, an orthogonal field of electrostatic field and static magnetic field is adopted to separate the electron and –H, and then they're accelerated along their own beam line to hit the target to produce x-rays or neutrons.

The static electric filed is produced by two concentric surfaces as the electrodes, and the magnetic poles, composed of two parallel permanent magnets, produce static magnetic field, and those two pairs of poles are parallel with the moving direction of the particle, as Fig2 shown. The motion of electric particle in the orthogonal field is talked in detail in the theory of energy selector [18]. Similarly in the particle separate case, as Fig3 shown, the forces on the electrical particle, briefly speaking, are:

The electrostatic field force: $\vec{F}_e = \vec{E}q$    (1)

The static magnetic field force: $\vec{F}_m = q\vec{v} \times \vec{B}$    (2)

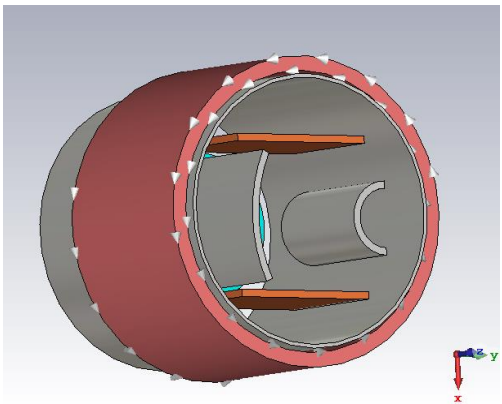
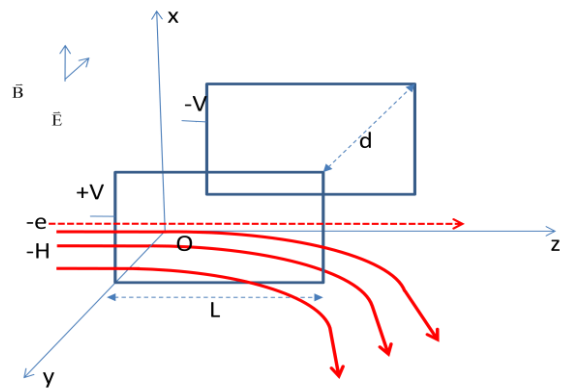

Fig2. Structrure of separate apparatus: two concentric surface as electric pole and two parallel plane as magnetic pole. A focus

Fig3. Motion of particles in orthognoal field: With electricfield opposite to y direction, and magneticfield in x dirction, and particles travel coil is outside of an electric shield.　　　　　in z dirction.

And, as for the non-relativistic particles, the relationship between the particle energy and velocity satisfies the energy equation, (3):

$$E = \frac{1}{2}mv^2 \quad (3)$$

From (2)&(3), the relationship between the magnetic force and particle energy satisfies Eq.(4), with the premise of $\vec{B} \perp \vec{v}$,

$$F_m = Bqv = Bq\sqrt{\frac{2E}{m}} \quad (4)$$

So, to the particle with same energy and charge, the velocity is inversely proportional to the half square root of mass, so does the magnetic force. In the separate zone, the electric force on the electron is designed to be balanced by the magnetic force, which means in y direction (5):

$$E_y = B_x v_e = B_x\sqrt{\frac{2E}{m_e}} \quad (5)$$

And because the mass of –H is much smaller than that of electron, according to eq. (4), the magnetic force on the –H is much smaller than that of electron, so that the magnetic force on –H is negligible when compared with that of electric force. In that ideal situation, the electron is travelling with uniform velocity, and the –H is moving similarly with horizontal projection. And as the result of different moving action, those two kinds of particles, with same energy but different mass, move with different trajectories, just as Fig3 shown.

In fact, because of the space charge effect and the existence of transverse speed, the beam will diverge transversely, which is especially obvious to the lower energy electric particle, so an axial magnetic field realized by a focus coil is adopt to restraint the transverse motion of electron. As for –H, a designation of a curved surface of electrical pole, Fig2, instead of a plane pole, Fig3, is adopted to give a concentric force to repress the transverse dispersion.

## 4 Designation of electrode

The electrode of the binocular structure composes of barrier node, coaxial node, and three accelerate node (Acc1, Acc2, Acc3). The barrier node is used to form a potential mirror to reflect the -H; the coaxial node is used to diminish the scale of the beam; the accelerate node is to accelerate the particle to the final energy. With the shape of pierce electron gun, which is wildly applied in static electron emission equipment, those three electrodes can also restrain the emittance of the beam more or less [19-20].

As the deviation of –H is realized by a transverse electric force, there is a transverse component of velocity, which is hard to be diminished by the pure accelerating field along the beam line. By importing a negative electric pole to form a transverse potential barrier, as Fig4 shown, the –H is reflected to the original direction, however, the emittance of -H gets worse. To improve the beam quality factor of -H, a coaxial electric node is adopted just inner the first accelerate node. With a lower potential than that of the first accelerate node, a sunken potential well forms inner the first accelerate node. With the centrifugal field inner the two sides of the coaxial node, the –H bears a centripetal force when passing through the coaxial electric node.

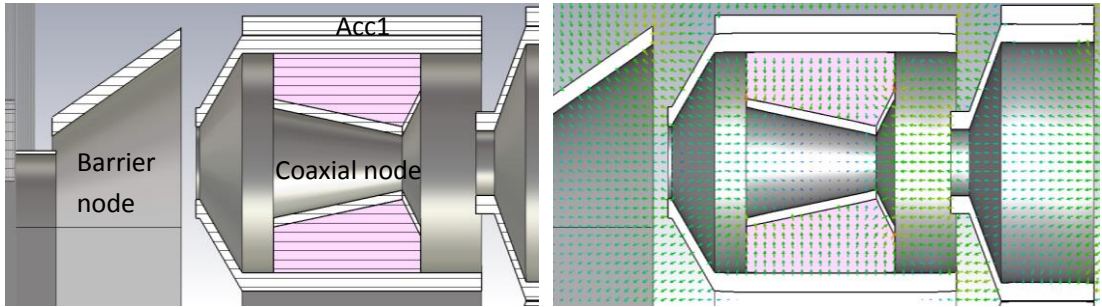

Fig4. Sectional view of barrier and accelerate node, and the electrical field distribution around the Acc1 node. The coaxial node is also coaxial with the Acc1 node of –H, separated and settled by ceramic.

5 Result and discussion

Based on the guidance mentioned above, a numerical designation and simulation of the binocular structure is carried out, with the purpose of verifying the feasibility of this designation. The energy of the particles entering the separate zone is around 1keV, as the result of the potential difference between the cathode and anode, which is a median energy of particle extracted from plasma [21-23].

Once the energy is settled, so does the numeric relationship between the electric field and magnetic field. Meanwhile, the deviation distance of the two beam line is decided by the transverse separation electric field, in this case, the distance between the two beam lines is settled to be 15mm. With the combination of electric field, the –H will first be accelerated transversely until out of the separate zone, and then drift out through the extracting hole, and then decelerated transversely by the potential barrier, forming a reverse writing 'z' trajectory, as Fig5 shown. Because of the non-uniform distribution of potential barrier, to particles with the same velocity entering the separate zone, the particles near the center need to move longer route to diminish the transverse velocity, which means the particles will gather together transversely. The angle and value of potential barrier affect the behavior of –H greatly, as Fig6 shown. Through an optimization, a structure of 35° and -500V is chosen.

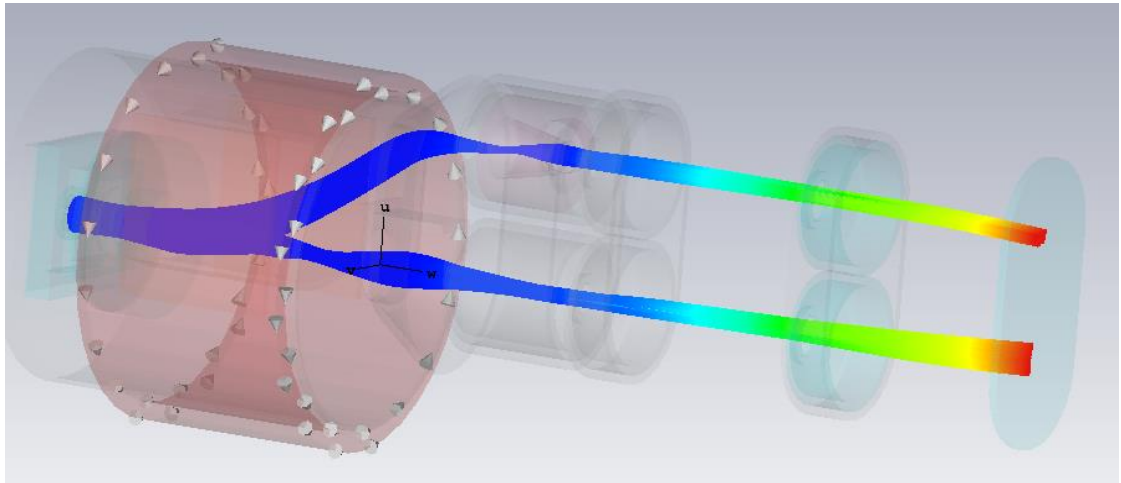

Fig5. Trajectory of the particles in the binocular structure, with electron trajectory below, and –H trajectory above.

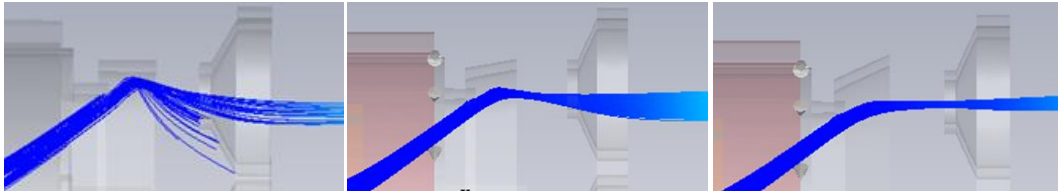

Fig6. Trajectory of –H with different potential barrier

The three accelerating electrodes are simply three sets of key parameters, such as the diameter of beam holes, the pierce angle and pole length, as table1 shown. As Fig5 shown, the –H gathers together transversely soon after the first node, and then should scatter quickly after the gathering point. With the help of the accelerate potential and the potential well of the coaxial node near the gathering point, the transverse motion is repressed, which proved to play an important role in the control of the beam quality of -H. By adjusting the structure parameters and potential combination of the accelerating electrodes, the trajectory of particles, with proper beam scale, is successfully corrected to their own beam line.

Table1. Parameters of electrodes

|  | Radius (in/mm) | Radius (out/mm) | Potential (kV) | Angle (°) | Position (mm) | Length (mm) |
|---|---|---|---|---|---|---|
| Acc1 | 2 | 6 | 7 | 30 | 48.8 | 15.5 |
| coaxial node | 3 | 4.5 | 1.5 | - | 53 | 8 |
| Acc2 | 1.8 | 6.5 | 12 | 30 | 64 | 7.7 |
| Acc3 | 2.3 | 6.5 | 87 | 25 | 92 | 7 |

Fig7 and Fig8 give the information of particle transportation and beam shape respectively. As Fig7 shown, the total number of electron and –H that hit target is almost the same as that of particle entering the separation zone, which means there is almost no loss of particle during the transportation after the separate zone. And the transverse distribution of particles are also adjusted to a size smaller than 3.5mm, which is smaller than that of beam near the source and in separate zone, just as Fig8 shows.

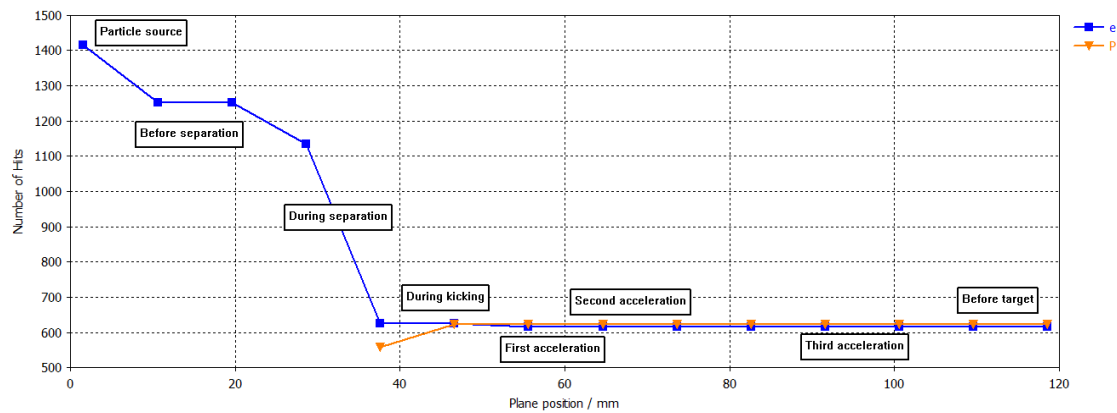

Fig7. Number counter of beam transportation, with blue 'e' represent monitor along the electron beam line, and orange 'P' represent monitor 14mm away from the electron beam line to monitor -H. The total hits of particle at the last monitor plan is almost the same as that of particle before separation.

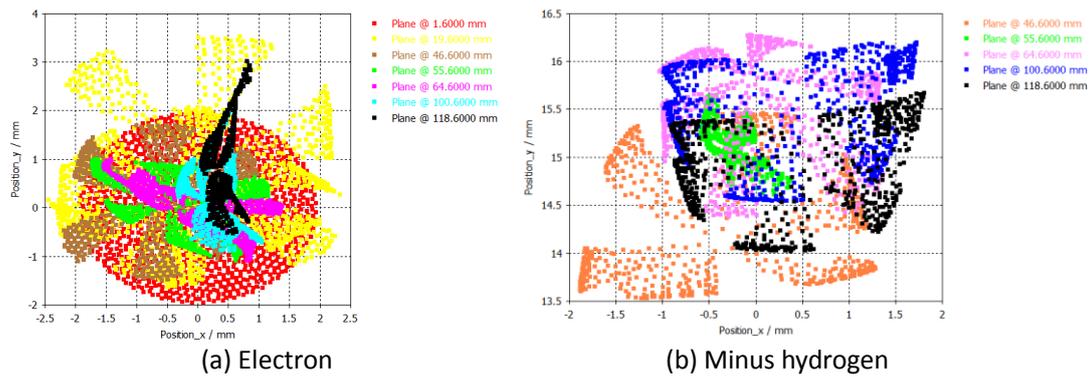

(a) Electron  (b) Minus hydrogen

Fig8. (a) Electron transverse distribution of different monitor plan. (b) –H transverse distribution of different monitor plan after separate zone. Red dots are particles just emitted from the cathode, with a complete circle. Yellow dots, brown dots, green dots, purple dots, blue dots, and black dots are particles inner the separate zone, after kicking zone, in Acc1, in Acc2, after Acc3, and before target. The electrons quickly gather together in x direction, but disperse in y direction soon after the kicking zone, and form a dot with a size of 1mm*3.5mm at last. The –H gathers together under the function of coaxial node inner Acc1 and form a dot with a size around 2.75mm*1.5mm. Both of the size is smaller than that of the particle source 4mm.

6. Conclusion

Under the guidance of the key rule of energy selector and potential combination, a binocular structure, used to produce complex rays by electron and –H, has been designed, with a pierce gun, instead of a plasma source, as the particle source. The simulation result shows that the combination of the two technique proves to be useful and effective. The two kinds of particle, with same energy and charge, are separated successfully, and lossless transported to the target with a rectangle spot of 1mm*3.5mm for electron, which can be tuned by the focus coil, and 2.75mm*1.75mm for –H, which can be tuned by coaxial node potential. Even though there're still lots of work need to do in detail before a verification on a figurative structure can be taken out, this design gives a way and feasible suggestion to realize a complex source of electron and -H. And at last, thanks the suggestion and help from team members during the work.